\documentclass[suppldata]{interact}
\usepackage{enumitem}
\newcounter{Bx}
\usepackage{epstopdf}
\usepackage[caption=false]{subfig}

\usepackage[numbers,sort&compress]{natbib}
\bibpunct[, ]{[}{]}{,}{n}{,}{,}
\renewcommand\bibfont{\fontsize{10}{12}\selectfont}
\usepackage{enumitem}
\theoremstyle{plain}

\theoremstyle{definition}

\theoremstyle{remark}

\newtheorem{theoremA}{Theorem}

\newcommand{\itemB}{%
    \addtocounter{Bx}{1}
    \item[B\theBx.]}
\begin{document}

\title{Wavelet Based Periodic Autoregressive Moving Average Models}

\author{
\name{Rhea Davis\textsuperscript{a}\thanks{CONTACT Rhea Davis. Email: rheadavisc@gmail.com} and N. Balakrishna\textsuperscript{a,b}}
\affil{\textsuperscript{a}Department of Statistics, Cochin University of Science and Technology, Kochi,  India; \textsuperscript{b}Department of Mathematics and Statistics, Indian Institute of Technology, Tirupati, India}
}

\maketitle

\begin{abstract}
This paper proposes a wavelet-based method for analysing periodic autoregressive moving average (PARMA) time series. Even though Fourier analysis provides an effective method for analysing periodic time series, it requires the estimation of a large number of Fourier parameters when the PARMA parameters do not vary smoothly. The wavelet-based analysis helps us to obtain a parsimonious model with a reduced number of parameters. We have illustrated this with simulated and actual data sets. 
\end{abstract}


\section{Introduction}
Time series with seasonal or periodic patterns occur in many fields of applications such as economics, finance, demography, astronomy, meteorology, etc. The commonly used seasonal autoregressive integrated moving average (SARIMA) models may not capture some of the important local features of such time series (see \cite{lund2000recursive}). An alternative approach suggested is to model these data by periodically stationary time series. A process $\{\Tilde{Y}_t\}$ is said to be periodically (weak) stationary if mean function $\mu_t = E(\Tilde{Y}_t)$ and autocovariance function $\gamma_t(h) = Cov(\Tilde{Y}_t, \Tilde{Y}_{t+h}), h \in \mathbb{Z}$, are periodic functions with same period $\nu$ (see \cite{pagano1978periodic}). A particular class of models having periodic stationarity is the periodic autoregressive moving average models of period $\nu$ having autoregressive parameter $p$ and moving average parameter $q$ [$\text{PARMA}_{\nu}(p,q)$], given in  \cite{anderson1993asymptotic} is 
\begin{equation}\label{PARMAeq}
    Y_t - \sum_{j=1}^{p}\phi_t(j)Y_{t-j} = \varepsilon_t-\sum_{j=1}^{q}\theta_t(j)\varepsilon_{t-j}, t=1,2,\dotsc, 
\end{equation}
where (i) $Y_t = \tilde{Y}_t-\mu_t, \mu_t = \mu_{t+k\nu}, k \in \mathbb{Z},$ (ii) $\phi_t = \phi_{t+k\nu}, k \in \mathbb{Z}$, (iii) $\theta_t = \theta_{t+k\nu}, k \in \mathbb{Z}$, and (iv) $\sigma^2_t = \sigma^2_{t+k\nu}, k \in \mathbb{Z}$, where $\sigma^2_t$ is the variance of the mean zero random sequence $\{\varepsilon_t\}$. Thus, $\{\delta_t\} = \{\sigma_t^{-1} \varepsilon_t\}$ is an iid sequence of mean 0 and variance 1 random variables. Note that PARMA reduces to ARMA when $\nu = 1$. PARMA models have been applied in fields as diverse as economics (\cite{parzen1979approach} and \cite{franses2004periodic}), climatology (\cite{jones1967time}, \cite{bloomfield1994periodic} and \cite{anderson2021parsimonious}), signal processing \cite{gardner1975characterization} and hydrology (\cite{vecchia1985periodic}, \cite{vecchia1985maximum}, \cite{anderson1993asymptotic}, \cite{anderson1998modeling} and  \cite{anderson2007fourier}). However, PARMA models contain a large number of parameters. For example, for a monthly series of period $\nu = 12$, the total number of parameters to be estimated is $12 \times 4 = 48$. Often, the inclusion of a large number of parameters leads to overfitting. Hence, it is important to develop methods that result in parsimonious models having less number of parameters.

This article is organised in the following manner: In Section 2, the estimators of the PARMA model and their several properties are mentioned. The Fourier-PARMA model is presented in Section 3. Section 4 gives a brief introduction to wavelet analysis with a special focus on discrete wavelet transform. In Section 5, our proposed wavelet-PARMA model is expounded. Section 6 consists of a simulation study of our proposed wavelet-PARMA model. The applicability of the proposed model is demonstrated using real data in Section 7. Section 8 contains the conclusion of our study. We have utilised the results from various articles, and for easy reference, we have included them in the Appendices.

\section{Estimation of PARMA parameters}

It is assumed hereafter that model (\ref{PARMAeq}) admits a causal representation (see \cite{anderson2021parsimonious})
$$
Y_t=\sum_{j=0}^{\infty} \psi_t(j) \varepsilon_{t-j},
$$
where $\psi_t(0)=1$ and $\sum_{j=0}^{\infty}\left|\psi_t(j)\right|<\infty,$ for all $t$. Note that $\psi_t(j)=\psi_{t+k v}(j)$ for all $j$. Also, $Y_t=\tilde{Y}_t-\mu_t$ and $\varepsilon_t=\sigma_t \delta_t$ where $\left\{\delta_t\right\}$ is i.i.d.

Given data $\tilde{Y}_0,\tilde{Y}_1, \dotsc ,\tilde{Y}_{N\nu-1}$, where $N$ is the number of cycles, the vector $\boldsymbol{\mu} = \{\mu_0, \dotsc , \mu_{\nu-1}\}^{\prime}$ is estimated using
\begin{equation}\label{mu.est}
    \hat{\mu}_i = \frac{1}{N}\sum_{j=0}^{N-1}\tilde{Y}_{j\nu+i}, \,\,  i = 0,1, \dotsc ,\nu-1. 
\end{equation}
The sample autocovariance for lag $m$ and season $i$ is estimated using
\begin{equation}\label{gamma.est}
    \hat{\gamma}_i(m) = \frac{1}{N}\sum_{j=0}^{N-1}(\tilde{Y}_{j\nu+i}-\hat{\mu}_i)(\tilde{Y}_{j\nu+i+m}-\hat{\mu}_{i+m}),\,\,  i = 0,1, \dotsc ,\nu-1, 
\end{equation}
where $\hat{\mu}_i$ is estimated using (\ref{mu.est}) and whenever $t>N\nu-1$, the corresponding term is set as zero. Let
\begin{equation*}
    \hat{Y}_{i+n} = \sum_{j=1}^n \theta_{n,j}^{(i)}(Y_{i+n-j}-\hat{Y}_{i+n-j}),\, n \geq 1,
\end{equation*}
denote the one-step predictor of $Y_{i+n}$ that minimises the mean square error, $v_{n,i} = E(Y_{i+n}-\hat{Y}_{i+n})^2$. The estimates of $\theta_{n,j}^{(i)}$ and $v_{n,i}$ can be obtained by substituting the estimates of autocovariance function (\ref{gamma.est}) in the recursive relation of innovation algorithm for periodic stationary processes given in Theorem \ref{A.innovationalgstat}. Then, Theorem \ref{A.psi theorem} and Theorem \ref{A.con.sigma2} show that $\hat{\theta}_{n,j}^{\langle i-n\rangle}$ and $\hat{v}_{n,\langle i-n \rangle}$ are consistent estimators for $\psi_i(j),$ for all $j$, and $\sigma_i^2$ respectively. Here the notation, $\langle b \rangle$ is $b-\nu \lfloor b/{\nu} \rfloor \, \text{for} \, b = 0, 1, \dotsc$ and $\nu+b-\nu\lfloor b/{\nu} + 1 \rfloor$ for $b = -1, -2, \dotsc$, where $\lfloor \cdot \rfloor$ is the greatest integer function.

To demonstrate our proposed ideas we are focusing particularly on two classes of PARMA models, described in the following subsections.
\subsection{Periodic autoregressive model of order 1 (PAR(1))}
Consider the $\text{PAR}_{\nu}(1)$ model given by,
\begin{equation}\label{particularPAR}
    Y_t = \phi_tY_{t-1}+\varepsilon_t,
\end{equation}
where (i) $Y_t = \tilde{Y}_t - \mu_t, \mu_t = \mu_{t+k\nu}, k \in \mathbb{Z},$ (ii) $\phi_t = \phi_{t+k\nu}, k \in \mathbb{Z}$ and (iii) $\sigma_t^2 = \sigma_{t+k\nu}^2, k \in \mathbb{Z}$ where $\sigma_t^2$ is the variance of the mean zero normal random sequence $\{\varepsilon_t\}$. 
For the model (\ref{particularPAR}), Theorem \ref{A.phi.equal.psi.result} tells us that $\{\psi_0(1),\psi_1(1),\dotsc,\psi_{\nu-1}(1)\}^{\prime} = \{\phi_0,\phi_1,\dotsc,\phi_{\nu-1}\}^{\prime}$. Thus, $\hat{\boldsymbol{\phi}} = \{\hat{\phi}_0,\hat{\phi}_1,\dotsc,\hat{\phi}_{\nu-1}\}^{\prime}$ can be obtained using the consistent estimates of $\psi_i(1), i = 0, \dotsc, \nu-1.$ Moreover, Theorem \ref{A.asym.dist.phi} tells us that the estimator of $\boldsymbol{\phi}$ has asymptotic normality. Also, $\hat{\boldsymbol{\mu}}$ and $\hat{\boldsymbol{\sigma}} = \{\hat{v}_{n,\langle 0-n \rangle}, \hat{v}_{n,\langle 1-n \rangle}, \dotsc, \hat{v}_{n,\langle \nu-1-n \rangle}\}^{\prime}$ have asymptotic normality as stated by Theorem \ref{A.asymp.distbn.mu} and Theorem \ref{A.asymp.distb.var}, respectively. 

\subsection{PARMA(1,1)}
Consider the $\text{PARMA}_{\nu}(1,1)$ given by,
\begin{equation}\label{particularPARMA}
    \tilde{Y}_t = \phi_t\tilde{Y}_{t-1} + \varepsilon_t - \theta_t\varepsilon_{t-1},
\end{equation}
where (i) $\phi_t = \phi_{t+k\nu}, k \in \mathbb{Z}$ (ii) $\theta_t = \theta_{t+k\nu}, k \in \mathbb{Z}$ and $\{\varepsilon_t\}$ is a sequence of normal random variables with mean 0 and variance 1.
For the model (\ref{particularPARMA}), Theorem \ref{A.PARMA.estimates} tells us that, 
\begin{equation}
    \phi_t = \frac{\psi_t(2)}{\psi_{t-1}(1)}.
\end{equation}
\begin{equation}
    \theta_t = \phi_t - \psi_t(1).
\end{equation}
Thus, $\hat{\phi}_t$ and $\hat{\theta}_t$ are obtained by replacing $\psi_t$ with its corresponding consistent estimate. The estimator of $\boldsymbol{\phi}$ has asymptotic normality as given in Theorem \ref{A.Q.matrix.thm}. Also, Theorem \ref{A.S.matrix.thm} tells us that $\hat{\boldsymbol{\theta}} = \{\hat{\theta}_1, \hat{\theta}_2, \dotsc, \hat{\theta}_{\nu-1}\}^{\prime}$ also has asymptotic normality. 

After estimating the PARMA parameters our objective is to reduce the number of parameters so that the final model is still able to capture the dependency structure adequately. Since parameter functions are periodic, it is natural to consider Fourier techniques to meet the objectives.

\section{Fourier-PARMA model}
Often, it is found that representing a function in terms of basis terms helps us to get new insights about the function, which is otherwise obscure. The selection of the set of basis functions depends on the kind of information we are interested in. The analysis of periodic functions usually involves the identification of prominent frequencies present in the signal. For this the function is represented as an infinite series of sinusoids, known as Fourier series, which is possible for almost all functions encountered in practice. The magnitude of the coefficients associated with sines and cosines indicates the strength of the corresponding frequency component present in the process.  

For a given vector $\mathbf{X}$, from \cite{anderson2021parsimonious},  the periodic function $\mathbf{X} = [X_t: 0 \leq t \leq \nu - 1]$ has the representation
\begin{equation} \label{DFT}
    X_t = c_{0}+\sum_{r=1}^{\ell}\left\{c_{r}\,cos\left(\frac{2\pi rt}{\nu}\right)+s_{r}\,sin\left(\frac{2\pi r t}{\nu}\right)\right\},
\end{equation}
where $c_{r}$ and $s_{r}$ are Fourier coefficients and 
\begin{equation*}
    \ell = \begin{cases}
        \nu /2, \quad \quad \quad \, \,\, \text{if} \, \nu \, even \\
        (\nu -1)/2, \quad \text{if}\,  \nu \, odd
    \end{cases}.
\end{equation*}
For convenience, (\ref{DFT}) is written as, 
\begin{equation}
    X_t = c_{0}+\sum_{r=1}^{\ell}\left\{c_{r}\,cos^*(r)+s_{r}\,sin^*(r))\right\},
\end{equation}
where $cos^*(r) = cos\left(\frac{2\pi rt}{\nu}\right)$ and $sin^*(r) = sin\left(\frac{2\pi rt}{\nu}\right)$.\\
\hfill\\
Let $\textbf{f} = \begin{cases}
    [c_{0},c_{1},s_{1},\dotsc,c_{(\nu-1)/2},s_{(\nu-1)/2}]^{\prime} \quad (\nu \, \text{odd}) \\
    [c_{0},c_{1},s_{1},\dotsc,s_{(\nu /2 -1)},c_{(\nu /2)}]^{\prime} \quad \quad (\nu \, \text{even})
\end{cases}$. \\
\hfill\\
Then, we have, from \cite{anderson2021parsimonious},
\begin{equation}\label{fx exp}
\textbf{f} = \mathcal{L}\mathcal{P}\mathcal{U}\mathbf{X},
\end{equation}
where $\mathcal{L}, \mathcal{P}$ and $\mathcal{U}$ are respectively given in equations (\ref{A.L.exp}), (\ref{A.P.exp}) and (\ref{A.U.exp}) of Theorem \ref{A.LPU.rep}.
$\mathcal{U}$ and $\mathcal{P}$ are unitary matrices. Using (\ref{fx exp}), we have, $\hat{\textbf{f}} = \mathcal{L}\mathcal{P}\mathcal{U}\hat{\mathbf{X}}$. Note that $\textbf{X}$ can be retrieved from $\textbf{f}$ by, 
\begin{equation*}
    \textbf{X} = \tilde{\mathcal{U}}^{\prime}\tilde{\mathcal{P}}^{\prime}\mathcal{L}^{-1},
\end{equation*}
where $\tilde{\mathcal{A}}^{\prime}$ denotes the conjugate transpose of matrix $\mathcal{A}$. Thus, $\textbf{X}$ and $\textbf{f}$ are representations of the same entity. Hence, to obtain a parsimonious PARMA model it is enough to represent the model using a reduced number of Fourier coefficients. Since it is observed that PARMA parameters often vary smoothly over time, \cite{anderson1993asymptotic}, \cite{anderson2007fourier}, \cite{tesfaye2011asymptotic} and \cite{anderson2021parsimonious} identified significant Fourier coefficients via a hypothesis test based on the asymptotic distributions of PARMA estimators. The significant Fourier coefficients are retained as it is while the insignificant coefficients are reduced to zero, thereby resulting in a parsimonious Fourier-PARMA model.

Although some success has been found in employing Fourier techniques they often fail when bursts and other transient events take place, which is often the case in many fields like economics and climatology. This is because, in the case of temporal events, all the Fourier coefficients get unduly affected as all the observations are involved in the computation of the coefficients.

\section{Wavelet Analysis}

Wavelet techniques are an excellent alternative to Fourier methods. Some applications of wavelets in statistics include estimation of functions like density, regression, and spectrum, analysis of long memory process, data compression, decorrelation, and detection of structural breaks. A “wavelet” means small wave (see \cite{pewav2000}, \cite{nason2008wavelet} and \cite{hardle2012wavelets}). This essentially means that a wavelet has oscillations like a wave. However, it lasts only for a short duration unlike sine wave, which spans infinitely to both sides of the time axis. They are constructed so that they possess certain desirable properties. Many different wavelet functions and transforms are available in the literature. However, here we are considering only the orthogonal discrete wavelet transform (DWT) since the approximation based on an orthogonal set is best in terms of mean square error, as indicated by projection theorem (see \cite{brockwell1991time}).

Fourier methods are enough when the frequencies of the signal do not change over time. However, Fourier techniques fail when different frequencies occur at different parts of the time axis. Hence, it is imperative to find techniques that have both time resolution and frequency resolution. But, by Heisenberg's Principle, it is impossible to identify time and frequency simultaneously with arbitrary precision (see \cite{mallat1999wavelet} and  \cite{vidakovic1999statistical}) as there is a lower bound for the error occurred. However, wavelets intelligently bypass this difficulty by using short time windows to capture high frequency, so that we have good time resolution, and wider time windows for low frequencies so that we have good frequency resolution. Thus, wavelets can pick out characteristics that are local in time and frequency. Wavelets are functions of scale, instead of frequency. A scale can be loosely interpreted as a quantity inversely proportional to frequency. This makes the wavelet transform an exceptional tool for studying non-stationary signals. For the latest works on the applications of wavelets to time series, one can refer to \cite{mcgonigle2022trend}, \cite{wang2022multiscale}, and \cite{mohammadi2023using}. 

Let $\mathbf{X}$ be a vector of size $N = 2^J, J \, \in \, \mathbb{Z}^+$, $\mathcal{W}$ is an $N \times N$ orthogonal matrix used for performing discrete wavelet transform (DWT), then the DWT of $\mathbf{X}$ is given by,
\begin{equation}
    \mathbf{W} = \mathcal{W}\mathbf{X} = \begin{bmatrix}
    {V}_{J}\\ 
    \mathbf{W}_{1}\\
    \vdots\\
    \mathbf{W}_{J}
    \end{bmatrix} = 
    \begin{bmatrix}
        W_0\\
        W_1\\
        \vdots\\
        W_N
    \end{bmatrix}
\end{equation}
Here, ${V}_{J} = W_0$, is the scaling coefficient that is proportional to the average of all the data, i.e, 
\begin{equation}\label{scaling coefficient}
    {V}_{J} = W_0 =  k_1*\frac{1}{N}\sum_{i=1}^{N}X_i,
\end{equation}
where the proportionality constant $k_1$ depends on the particular wavelet employed. $\mathbf{W}_{j}, j \in \{1,2,\dotsc,J\}$ contains $2^{j-1}$ wavelet coefficients associated with changes on a scale $N/2^{j-1}$.  
For example, the Haar DWT of $\mathbf{X} = [X_0,\dotsc,X_{15}]^{\prime}$ is given by,
\begin{equation}\label{example}
    \begin{bmatrix}
    \frac{1}{4}(X_{15}+ \dotsc + X_0)\\
    	\frac{1}{4}(-X_{15}+\dotsc-X_8+X_7-\dotsc + X_0)\\
      	\frac{1}{\sqrt{8}}(-X_7+\dotsc-X_4+X_3+\dotsc +X_0)\\
       \frac{1}{\sqrt{8}}(-X_{15}-\dotsc-X_{12}+X_{11}-\dotsc +X_8)\\
 \frac{1}{2}(-X_3-X_2+X_1+X_0)\\    
        	\vdots\\ 
         \frac{1}{2}(-X_{15}-X_{14}+X_{13}+X_{12})\\
         	\frac{1}{\sqrt{2}}(-X_1+X_0)\\
          \vdots\\
 \frac{1}{\sqrt{2}}(-X_{15}+X_{14})
\end{bmatrix}. 
\end{equation}
Note that the wavelet coefficients are localised in time. For example, the last wavelet coefficient only involves the observations $X_{14}$ and $X_{15}$. 
 Since $\mathcal{W}$ is an orthogonal matrix, the original vector $\mathbf{X}$ can easily be reconstructed via the inverse DWT
\begin{equation}\label{inverseDWT}
    \mathbf{X} = \mathcal{W}^T\mathbf{W},
\end{equation}
where $\mathcal{W}^T$ denotes the transpose of $\mathcal{W}$. Note that from (\ref{inverseDWT}), $\textbf{X}$ and $\textbf{W}$ are representations of the same quantity and hence it is enough to represent the parameters of the PARMA model in terms of discrete wavelet transform coefficients. 

Another interesting advantage of the DWT, especially pertaining to our objective of achieving parsimonious PARMA models, is the fact that DWT redistributes the energy or variability contained in a sequence (see \cite{vidakovic1999statistical}). Hence, the crux of the sequence, spread throughout the original sequence, is concentrated in a few wavelet coefficients (see \cite{ogden1997essential} and  \cite{vidakovic1999statistical}). Hence, we propose a wavelet-PARMA model that captures the dependency structure adequately based on the idea of retaining only a few significant wavelet coefficients. We identify the significant wavelet coefficients by developing a hypothesis test, following the parallel approach given in \cite{anderson2021parsimonious} for Fourier coefficients, utilizing the asymptotic distributions of the estimators of the PARMA parameters. 

\section{Wavelet - PARMA model}
Suppose random vector $\hat{\mathbf{X}}$ satisfies, 
\begin{equation}\label{asynormalX}
    N^{1/2}(\hat{\mathbf{X}}-\mathbf{X}) \Rightarrow \mathcal{N}(0,\Sigma).
\end{equation}
Using Theorem \ref{A.broc.asymnormaltheorem}, we have, 
\begin{equation}\label{distbn.of.W}
    N^{1/2}(\hat{\mathbf{W}} - \mathbf{W}) \Rightarrow \mathcal{N}(0,R_\textbf{W}),
\end{equation}
\begin{equation}\label{Reqnwavelet}
\text{where} \quad R_\textbf{W} = \mathcal{W}\Sigma\mathcal{W}^T.
\end{equation}
The main idea is to identify the statistically significant wavelet coefficients of the PARMA estimator vectors, using an appropriate test procedure so that the other coefficients can be nullified, thereby obtaining a parsimonious PARMA model. The null hypothesis of the test is $H_0: \mathbf{W} = (V_J,0,\dotsc,0) = (W_0,0,\dotsc,0)$. Under $H_0$, $\mathbf{X} = k_2*V_J = k_2*k_1*\frac{1}{N}\sum_{i=1}^NX_i = k*\frac{1}{N}\sum_{i=1}^NX_i$, where $k$ depends on the type of wavelet used. Since the parameter vectors have all elements the same under $H_0$, the PARMA process reduces to a stationary ARMA process with $\phi_t = \phi$ for all $t \in \{0,1,\dotsc,\nu-1\}$, and similarly for other parameter vectors. Thus, the null hypothesis states that the model is stationary. From (\ref{distbn.of.W}), we have, 
\begin{equation}
    N^{1/2}(\hat{W}_i - W_i) \Rightarrow \mathcal{N}(0,[{R_\textbf{W}}]_{ii}), \, \text{for} \, i = 1,2,\dotsc, \nu-1.
\end{equation}
Here, we will use the Bonferroni's test procedure (see \cite{howell2012statistical}) with $\alpha = 0.05$. In this case,  we wish to test the null hypothesis $H_0: |W_i| = 0$ versus $H_a:|W_i| \neq 0, i = 1,2,\dotsc,\nu-1$. The test statistics is 
\begin{equation}
    {Z_W}_i = \frac{\hat{W}_i}{\sqrt{[{R_\textbf{W}}]_{i,i}/N}}, i = 1,2,\dotsc \nu -1.
\end{equation}
Let $\alpha^* = \frac{\alpha}{\nu -1}$ and $Z_{\alpha^*/2}$ be the $\alpha^*/2$ tail quantile, i.e, $P(|Z|>Z_{\alpha^*/2}) = \alpha^*/2$, where $Z \sim \mathcal{N}(0,1)$. We reject the null hypothesis when $|{Z_W}_i|>Z_{{\alpha^*}/2}$. Whenever the null hypothesis is not rejected, the corresponding element $W_i$ is set as zero. Hopefully, this will result in a sparse $\mathbf{W}^{\prime}$ vector where most of the coefficients are 0, resulting in a parsimonious wavelet-PARMA model.
\subsection{PAR(1) under $H_0$}
For the model considered in (\ref{particularPAR}), under null hypothesis, 
$\gamma(h) = \nu^{-1}\sum_{m=0}^{\nu-1}\gamma_m(h)$ and $\phi(k) = \nu^{-1}\sum_{m=0}^{\nu-1} \phi_m(k)$.
 In this case, 
 the asymptotic variance-covariance matrix of $\hat{\boldsymbol{\mu}}, \Sigma_{\boldsymbol{\mu}}$, becomes (see \cite{anderson2021parsimonious}),  
\begin{equation}
    (\Sigma_{\boldsymbol{\mu}})_{ii,} = \gamma(0)\left[\frac{1+r}{1-r}\right], \quad \, 0\leq i \leq \nu - 1,
\end{equation}
 where $r=\phi^{\nu}$ and for $j>i$, 
 \begin{equation}
     (\Sigma_{\boldsymbol{\mu}})_{ij} = \gamma(0)\left[\frac{\phi^{j-i}+\phi^{\nu+i-j}}{1-r}\right], \quad 0 \leq i,j \leq \nu - 1.
 \end{equation}
Since $\Sigma_{\boldsymbol{\mu}}$ is a symmetric matrix, we also have the elements $(\Sigma_{\boldsymbol{\mu}})_{ij, i>j, 0 \leq i,j \leq \nu -1}$. For hypothesis testing, the elements of $\Sigma_{\boldsymbol{\mu}}$ are replaced with their estimates. 

Under $H_0$, the asymptotic variance-covariance matrix of $\hat{\boldsymbol{\sigma^2}}$, $\Sigma_{\boldsymbol{\sigma^2}}$ is given by (see \cite{anderson2021parsimonious}), 

\begin{align}
(\Sigma_{\boldsymbol{\sigma^2}})_{ii}=&2 \gamma(0)^2\left[\frac{1+\phi^{2v}}{1-r^2}\right]-2 \phi \times 2 \gamma(0)^2\left[\frac{\phi+\phi^{2 v-1}}{1-r^2}\right]\\
&+\phi^2 \gamma(0)^2\left[\phi^2+\frac{1+3 r^2}{1-r^2}\right], \quad 0 \leq i \leq \nu -1,
\end{align}
and if $j>i$, we have, 
\begin{align}
(\Sigma_{\boldsymbol{\sigma}^2})_{ij}= & 2 \gamma(0)^2\left[\frac{\phi^{2(j-i)}+\phi^{2(v+i-j)}}{1-r^2}\right]\\
&-2 \gamma(0)^2 \phi\left[\frac{\phi^{2 j-2 i-1}}{1-r^2}+\frac{\phi^{2 v-2 j+2 i+1}}{1-r^2}\right] \nonumber 
-2 \gamma(0)^2 \phi\left[\frac{\phi^{2 j-2 i+1}+\phi^{2 \nu-2 j+2 i-1}}{1-r^2}\right]\\
&+2 \gamma(0)^2 \phi^2\left[\frac{\phi^{2 j-2 i}+\phi^{2 v-2 j+2 i}}{1-r^2}\right], \quad 0 \leq i,j \leq \nu -1.
\end{align}
Since $\Sigma_{\boldsymbol{\sigma^2}}$ is a symmetric matrix, we also have the elements $s_{i,j},i>j.$ For hypothesis testing, the elements of $\Sigma_{\boldsymbol{\sigma^2}}$ are replaced with their estimates.

Under $H_0$, $\sigma_i^2 = \sigma^2,$ for all $i$ and so from Theorem \ref{A.asym.dist.phi}, we get that the asymptotic variance-covariance matrix of $\hat{\boldsymbol{\phi}}, \Sigma_{\boldsymbol{\phi}}$, as the identity matrix $\mathcal{I}$ of order $\nu$, i.e., $\Sigma_{\boldsymbol{\phi}} = \mathcal{I}$. 

\subsection{PARMA(1,1) under $H_0$}
For the model considered in (\ref{particularPARMA}), under null hypothesis, $\psi(j) = \nu^{-1}\sum_{t=0}^{\nu-1}\psi_t(j) ,j = 1,2, \phi = \nu^{-1}\sum_{t=0}^{\nu-1}\phi_t$ and $\theta = \nu^{-1}\sum_{t=0}^{\nu-1}\theta_t$. Under $H_0$, the asymptotic variance-covariance matrix of $\hat{\boldsymbol{\phi}}, \mathcal{Q}$, given in Theorem \ref{A.Q.matrix.thm}, reduces to 
\begin{equation}\label{Q}
    \mathcal{Q} = \sum_{k,l=1}^{2}\mathcal{H}_{\ell}\mathcal{V}_{\ell k}\mathcal{H}_{k}^{\prime},
\end{equation}
where  
\begin{align}
    \mathcal{V}_{11} &= \mathcal{I}, \label{V11}\\
    \mathcal{V}_{22} &= [\boldsymbol{\psi}^{\prime}(1)+1]\mathcal{I},\\
    \mathcal{V}_{12} &= \boldsymbol{\psi}(1)\Pi,\\
    \mathcal{V}_{21} &= \boldsymbol{\psi}(1)\Pi^{\prime}, \label{V21}\\
    \mathcal{H}_1 &= -\mathcal{F}_2\Pi^{-1}\mathcal{F}_1^{-2}.\\
    \mathcal{H}_2 &= \Pi^{-1}\mathcal{F}_1^{-1}\Pi, \\
    \mathcal{F}_j &= \boldsymbol{\psi}(j)\mathcal{I},\quad  \label{Fn} \\
    \boldsymbol{\psi}(j) &= \{\psi_0(j), \psi_1(j),\dotsc,\psi_{\nu - 1}(j)\}^{\prime},\\
   \text{and} \quad \Pi &= \begin{bmatrix}
0 & 1 & 0 & 0 & \dotsc &0 \\ 
0 & 0 & 1 & 0 & \dotsc & 0 \\ 
\vdots  &\vdots  &\vdots  &\vdots  &  &\vdots \\ 
 0& 0 & 0 & 0 & \dotsc & 1\\ 
 1& 0 & 0 & 0 & \dotsc & 0
 \label{Pi} \end{bmatrix}.
\end{align}
Also, the asymptotic variance-covariance matrix of $\hat{\boldsymbol{\theta}}, \mathcal{S},$ given in Theorem \ref{A.S.matrix.thm}, has the following form under $H_0$:
\begin{equation}
\mathcal{S} = \sum_{k,l=1}^{2}\mathcal{M}_{\ell}\mathcal{V}_{\ell k}\mathcal{M}_{k}^{\prime},
\end{equation}
where $\mathcal{M}_1 = -\mathcal{I}-\mathcal{F}_2\Pi^{-1}\mathcal{F}_1^{-2}$, $\mathcal{M}_2 = \Pi^{-1}\mathcal{F}_1^{-1}\Pi$, $\mathcal{V}_{lk}, 1\leq l,k \leq 2$ are given in equations (\ref{V11}) to (\ref{V21}) , $\mathcal{F}_j$ is given in (\ref{Fn}) and $\Pi$ is given in (\ref{Pi}).

If $\nu$ is not a power of 2, each of the estimator vectors is extended periodically to the nearest power of 2, say $\nu^{\prime}$. The corresponding asymptotic variance-covariance matrices are also extended periodically, resulting in matrices of order $\nu^{\prime}$.
\section{Simulation of wavelet-PARMA}
The observations are simulated from the $\text{PARMA}_{12}(1,1)$ model,
\begin{equation}
    \tilde{Y}_t = \phi_t\tilde{Y}_{t-1}+\varepsilon_t - \theta_t \varepsilon_{t-1},
\end{equation}
where (i) $\phi_t = \phi_{t+12k}$, (ii) $\theta_t = \theta_{t+12k}, k \in \mathbb{Z}$ and $\{\varepsilon_t\}$ is a sequence of normal random variables with mean 0 and variance 1. The parameter values and their corresponding estimates are given in Table \ref{wav PARMA true and estimates}.
\begin{table}
\tbl{True value and their estimates of PARMA model.}
{\begin{tabular}{|c|cccc|}
\hline
       & \multicolumn{4}{c|}{Parameter}                                                                   \\ \cline{2-5} 
       & \multicolumn{2}{c|}{$\phi$}                               & \multicolumn{2}{c|}{$\theta$}        \\ \hline
Season & \multicolumn{1}{c|}{True} & \multicolumn{1}{c|}{Estimate} & \multicolumn{1}{c|}{True} & Estimate \\ \hline
0      & \multicolumn{1}{c|}{0.67} & \multicolumn{1}{c|}{0.59}     & \multicolumn{1}{c|}{0.2}  & 0.06     \\
1      & \multicolumn{1}{c|}{0.7}  & \multicolumn{1}{c|}{0.73}     & \multicolumn{1}{c|}{0.23} & 0.29     \\
2      & \multicolumn{1}{c|}{0.69} & \multicolumn{1}{c|}{0.81}     & \multicolumn{1}{c|}{0.22} & 0.30     \\
3      & \multicolumn{1}{c|}{0.68} & \multicolumn{1}{c|}{0.77}     & \multicolumn{1}{c|}{0.21} & 0.24     \\
4      & \multicolumn{1}{c|}{0.67} & \multicolumn{1}{c|}{0.43}     & \multicolumn{1}{c|}{1.43} & 1.22     \\
5      & \multicolumn{1}{c|}{0.68} & \multicolumn{1}{c|}{0.72}     & \multicolumn{1}{c|}{1.44} & 1.46     \\
6      & \multicolumn{1}{c|}{0.69} & \multicolumn{1}{c|}{0.62}     & \multicolumn{1}{c|}{0.46} & 0.40     \\
7      & \multicolumn{1}{c|}{0.68} & \multicolumn{1}{c|}{1.04}     & \multicolumn{1}{c|}{0.47} & 0.81     \\
8      & \multicolumn{1}{c|}{1.83} & \multicolumn{1}{c|}{1.86}     & \multicolumn{1}{c|}{0.23} & 0.28     \\
9      & \multicolumn{1}{c|}{1.84} & \multicolumn{1}{c|}{1.83}     & \multicolumn{1}{c|}{0.24} & 0.25     \\
10     & \multicolumn{1}{c|}{0.53} & \multicolumn{1}{c|}{0.52}     & \multicolumn{1}{c|}{0.21} & 0.21     \\
11     & \multicolumn{1}{c|}{0.52} & \multicolumn{1}{c|}{0.68}     & \multicolumn{1}{c|}{0.23} & 0.37     \\ \hline
\end{tabular}}
\label{wav PARMA true and estimates}
\end{table}
Since period 12 is not a power of 2, each of the vectors has been extended periodically to the nearest power of 2, i.e, 16. We have taken the number of cycles, $N$ to be 500. 

\begin{figure}
    \centering \includegraphics{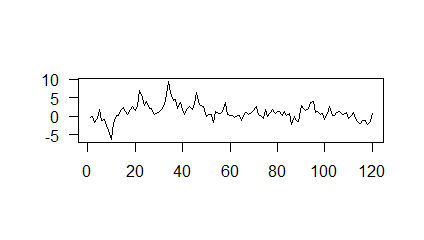}
    \caption{Plot of simulated observations}
    \label{wav PARMA sim 2 plot}
\end{figure}

\begin{figure}
\centering
\subfloat[ACF plot of PARMA(1,1) residuals.]{%
\resizebox*{5cm}{!}{\includegraphics{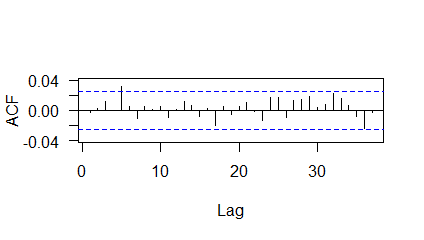}}}\hspace{5pt}
\subfloat[Histogram of PARMA (1,1) residuals superimposed with normal PDF.]{%
\resizebox*{5cm}{!}{\includegraphics{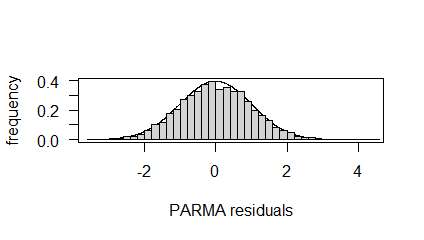}}}
\caption{PARMA(1,1) Residual Diagnostics} \label{PARMA sim 2 residual diagnostics}
\end{figure}

The plot of the generated observations is given in Figure \ref{wav PARMA sim 2 plot}.
7 iterations of the innovation algorithm was carried out to calculate the parameter estimates. 
The residuals are computed using the expression, 
\begin{equation}
    \hat{\epsilon}_t = \tilde{Y}_t - \hat{\phi}_t\tilde{Y}_{t-1}+\hat{\theta}_t \hat{\varepsilon}_{t-1},
\end{equation}
where the initial value $\hat{\varepsilon}_0$ is set as zero. The residual diagnostic plots are given in Figure \ref{PARMA sim 2 residual diagnostics}.
The Box Pierce test results are given in Table \ref{wav parma sim 2 box}.
\begin{table}
\tbl{Box-Pierce test for PARMA residuals}
{\begin{tabular}{ccccccc} \toprule
Lags    & 20     & 30     & 40     & 50    & 80     & 100    \\ \hline
p-value & 0.8844 & 0.8232 & 0.5431 & 0.706 & 0.8313 & 0.8898 \\ \bottomrule
\end{tabular}}
\label{wav parma sim 2 box}
\end{table}
The normality of the residuals was tested using Kolmogorov Smirnov test and it was found that the residuals are normally distributed (p-value = 0.8586). Thus, the residuals are uncorrelated and normally distributed. 

\begin{figure}
    \centering   \includegraphics{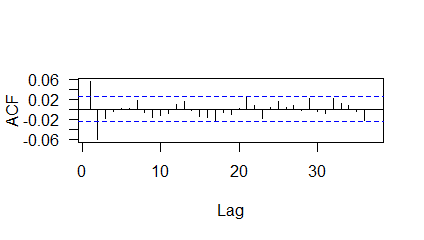}
    \caption{ACF plot of residuals of Fourier-PARMA model}   
    \label{ACF plot of residuals of Fourier-PARMA model sim 2}
\end{figure}

The obtained Fourier-PARMA model, by using the method outlined in \cite{anderson2021parsimonious} is given by,
\begin{equation*}
    \phi_t = 0.88-0.375\,sin^{*}(1)-0.27\,cos^*(2)+0.30\,sin^*(2)-0.195\,sin^*(4). 
\end{equation*}
\begin{equation*}
    \theta_t = 0.49-0.37\,cos^*(1)+0.195\,sin^*(1)-0.22\,sin^*(2)-0.25\,cos^*(4).
\end{equation*}
The ACF plot of residuals of the Fourier-PARMA model is given in Figure \ref{ACF plot of residuals of Fourier-PARMA model sim 2}.
The results of Box-Pierce test are given in Table \ref{parma sim 2 fourier box}.
\begin{table}
\tbl{Box-Pierce test for Fourier-PARMA model.}
{\begin{tabular}{ccccccc}
\toprule
Lags    & 20                     & 30                 & 40                    & 50     & 80      & 100    \\ \hline
p-value & $1.753 \times 10^{-5}$ & $7 \times 10^{-5}$ & $4.34 \times 10^{-5}$ & 0.0003 & 0.00299 & 0.0084 \\ \bottomrule
\end{tabular}}
\label{parma sim 2 fourier box}
\end{table}
Thus, the Fourier-PARMA model is not able to capture the dependencies among the observations adequately as is evident from the ACF plot given in Figure \ref{ACF plot of residuals of Fourier-PARMA model sim 2} and Box-Pierce test results given in Table \ref{parma sim 2 fourier box}.

\begin{figure}
    \centering    \includegraphics{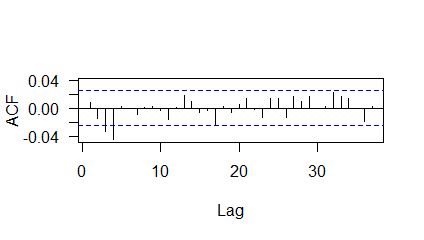}
    \caption{ACF plot of Wavelet-PARMA residuals}
    \label{wav parma sim 2 wav acf}
\end{figure}

\begin{table}
\tbl{DWT analysis of $\phi$ estimates and $\theta$ estimates of $\text{PARMA}_{12}(1,1)$. The statistically significant DWT coefficients ($|Z|>2.95$) are denoted by *. The coefficients considered in the Wavelet-PARMA model are indicated in bold.}
{\begin{tabular}{|c|cccc|}
\hline
   & \multicolumn{4}{c|}{Parameter}                                                                                       \\ \cline{2-5} 
   & \multicolumn{2}{c|}{$\phi$}                                         & \multicolumn{2}{c|}{$\theta$}                  \\ \cline{2-5} 
   & \multicolumn{1}{c|}{DWT coefficient} & \multicolumn{1}{c|}{Z-score} & \multicolumn{1}{c|}{DWT coefficient} & Z-score \\ \hline
0  & \multicolumn{1}{c|}{\textbf{3.37}}   & \multicolumn{1}{c|}{-}       & \multicolumn{1}{c|}{\textbf{1.69}}   & -       \\
1  & \multicolumn{1}{c|}{\textbf{-0.52*}}  & \multicolumn{1}{c|}{-5.30}   & \multicolumn{1}{c|}{\textbf{0.69*}}   & 6.12    \\
2  & \multicolumn{1}{c|}{0.03}            & \multicolumn{1}{c|}{0.21}    & \multicolumn{1}{c|}{\textbf{-1.06*}}  & -6.60   \\
3  & \multicolumn{1}{c|}{\textbf{0.71*}}   & \multicolumn{1}{c|}{5.09}    & \multicolumn{1}{c|}{0.08}            & 0.48    \\
4  & \multicolumn{1}{c|}{-0.13}           & \multicolumn{1}{c|}{-0.94}   & \multicolumn{1}{c|}{-0.10}           & -0.63   \\
5  & \multicolumn{1}{c|}{-0.26}           & \multicolumn{1}{c|}{-1.85}   & \multicolumn{1}{c|}{\textbf{0.73*}}   & 4.80    \\
6  & \multicolumn{1}{c|}{\textbf{1.25*}}   & \multicolumn{1}{c|}{8.97}    & \multicolumn{1}{c|}{-0.02}           & -0.16   \\
7  & \multicolumn{1}{c|}{-0.13}           & \multicolumn{1}{c|}{-0.94}   & \multicolumn{1}{c|}{-0.10}           & -0.63   \\
8  & \multicolumn{1}{c|}{-0.10}           & \multicolumn{1}{c|}{-0.72}   & \multicolumn{1}{c|}{-0.16}           & -1.21   \\
9  & \multicolumn{1}{c|}{0.03}            & \multicolumn{1}{c|}{0.21}    & \multicolumn{1}{c|}{0.05}            & 0.36    \\
10 & \multicolumn{1}{c|}{-0.21}           & \multicolumn{1}{c|}{-1.49}   & \multicolumn{1}{c|}{-0.17}           & -1.27   \\
11 & \multicolumn{1}{c|}{-0.30}           & \multicolumn{1}{c|}{-2.16}   & \multicolumn{1}{c|}{-0.29}           & -2.15   \\
12 & \multicolumn{1}{c|}{0.02}            & \multicolumn{1}{c|}{0.15}    & \multicolumn{1}{c|}{0.02}            & 0.14    \\
13 & \multicolumn{1}{c|}{-0.11}           & \multicolumn{1}{c|}{-0.80}   & \multicolumn{1}{c|}{-0.11}           & -0.85   \\
14 & \multicolumn{1}{c|}{-0.10}           & \multicolumn{1}{c|}{-0.72}   & \multicolumn{1}{c|}{-0.16}           & -1.21   \\
15 & \multicolumn{1}{c|}{0.03}            & \multicolumn{1}{c|}{0.21}    & \multicolumn{1}{c|}{0.05}            & 0.36    \\ \hline
\end{tabular}}
\label{PARMA wav sim2 results}
\end{table}
The discrete wavelet transform analysis was done using Haar wavelet. This wavelet was chosen for simulation because its structure is explicitly known. The results of Haar DWT analysis are given in Table \ref{PARMA wav sim2 results}. It is found that only 4 DWT coefficients of $\phi$ and 4 DWT coefficients of $\theta$ need to be incorporated into the wavelet-PAR model and so the final model contains only 8 parameters as opposed to 24 parameters in the original PARMA model. The ACF plot of the residuals of wavelet-PARMA model is given in Figure \ref{wav parma sim 2 wav acf}.
The results of Box-Pierce test are given in Table \ref{wav parma sim 2 wav box}.
\begin{table}
\tbl{Box-Pierce test for Wavelet-PARMA model.}
{\begin{tabular}{ccccccc}
\toprule
Lags    & 20     & 30     & 40     & 50     & 80     & 100    \\ \hline
p-value & 0.0889 & 0.1274 & 0.0931 & 0.1923 & 0.4974 & 0.6524 \\ \bottomrule
\end{tabular}}
\label{wav parma sim 2 wav box}
\end{table}
The residuals of the wavelet-PARMA model are uncorrelated. Hence, the wavelet-PARMA model with just 8 parameters is able to capture the dependency structure adequately. 
\section{Data Analysis}
\begin{figure}
    \centering  \includegraphics{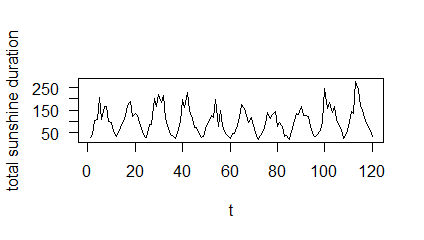}
    \caption{Plot of total sunshine duration in a month recorded by Ballypatrick station from 1966 to 1990.}
    \label{ballypatrick plot}
\end{figure}

\begin{figure}
\centering
\subfloat[ACF plot of PAR(1) residuals.]{%
\resizebox*{5cm}{!}{\includegraphics{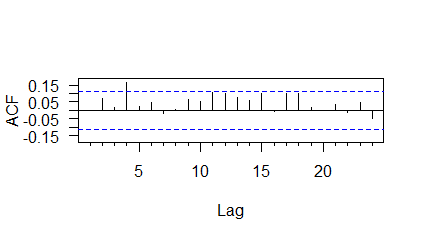}}}\hspace{5pt}
\subfloat[Histogram of PAR(1) residuals superimposed with normal PDF.]{%
\resizebox*{5cm}{!}{\includegraphics{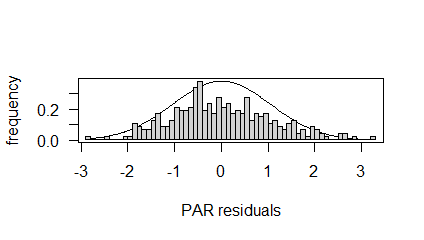}}}
\caption{PAR(1) Residual Diagnostics} \label{ballypatrick PAR res diag}
\end{figure}

The UK Meteorological (MET) Office, established in 1854, is the government body commissioned to analyse weather in the United Kingdom. They use their findings for forecasting and for issuing warnings. They offer data from 36 stations in the UK. For our analysis, we have considered the total sunshine duration in a month recorded in hours by the Ballypatrick weather station for the years 1966-1990. This data is available from the Kaggle site \textit{https://www.kaggle.com/datasets/josephw20/uk-met-office-weather-data/code}, where the total sunshine duration is given under the variable \textit{sun}. The data considered is fit using the $\text{PAR}_{12}(1)$ model given in equation (\ref{particularPAR}). Here, period $\nu = 12$ and the number of years (cycles), $N = 25$. The plot of the first 10 years is given in Figure \ref{ballypatrick plot}. It was found that 2 iterations of the innovation algorithm were enough to capture the model dynamics. The residuals of the PAR model was computed using, 
\begin{equation}\label{PAR residual expression}
    \hat{\delta}_t = \frac{\hat{Y}_t-\hat{\phi}_t\hat{Y}_{t-1}}{\hat{\sigma}_t},
\end{equation}
where $\hat{Y}_t = \tilde{Y}_t - \hat{\mu}_t$.
The PAR model diagnostics plots are given in Figure \ref{ballypatrick PAR res diag}. The results of the Box-Pierce test are given in Table \ref{ballypatrick acf box}. The normality of the residuals was tested using the Kolmogorov-Smirnov test and it was found that the residuals were normally distributed (p-value = 0.319). Thus, the residuals are uncorrelated and normally distributed. 

\begin{figure}
    \centering
\includegraphics{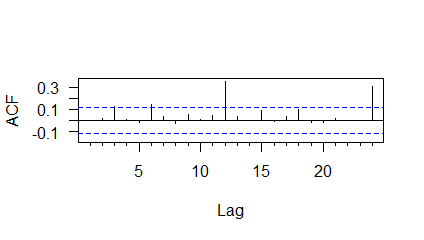}
    \caption{ACF plot of Fourier-PAR residuals.}
    \label{ballypatrick fourier acf}
\end{figure}

\begin{table}
\tbl{Box-Pierce test for PAR residuals}
{\begin{tabular}{ccc}
\toprule
Lags    & 20     & 30      \\ \hline
p-value & 0.0528 & 0.0601 \\ 
\bottomrule
\end{tabular}}
\label{ballypatrick acf box}
\end{table}
The obtained Fourier-PAR model, by using the method outlined in \cite{anderson2021parsimonious} is given by,
\begin{align*}
    \mu_t &= 106.45-60.67\,cos^*(1)+35.29\,sin^*(1)-12.02\,cos^*(2)+7.75\,cos^*(3)+6.88\,sin^*(4).\\
    \phi_t &= 0.09.\\
    \sigma^2_t &= 713.78-729.27\,cos^*(1)+395.27\,sin^*(1).
\end{align*}
The ACF plot of Fourier-PAR residuals is given in Figure \ref{ballypatrick fourier acf}.
The results of Box-Pierce test of Fourier-PAR residuals are given in Table \ref{ballypatrick fourier box}.
\begin{table}
\tbl{Box-Pierce test of Fourier-PAR residuals.}
{\begin{tabular}{ccc}
\toprule
Lags    & 20      & 30       \\ \hline
p-value & $1.3 \times 10^{-5}$ & $6.76 \times 10^{-8}$ \\ \bottomrule
\end{tabular}}
\label{ballypatrick fourier box}
\end{table}
Thus, it is clear that the Fourier-PAR model has not been able to capture the dependency structure adequately. 

\begin{figure}
    \centering
\includegraphics{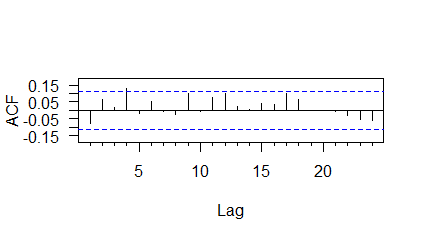}
    \caption{ACF plot of wavelet-PAR residuals.}
    \label{ballypatrick wav acf}
\end{figure}

The discrete wavelet transform analysis was done using Daubechies least asymmetric wavelet with 7 vanishing moments. This wavelet was chosen because the resulting wavelet-PAR model could capture the dependency structure adequately. The LA(7) DWT results are given in Table \ref{ballypatrick wav resuts}. The ACF plot of wavelet-PAR residuals is given in Figure \ref{ballypatrick wav acf}. The Box-Pierce test of wavelet-PAR residuals is given in Table \ref{ballypatrick wav box}. From these results, it is clear that the resulting wavelet-PAR model was able to capture the dependency structure adequately.  It is found that only 7 DWT coefficients of $\boldsymbol{\mu}$, 1 DWT coefficient of $\boldsymbol{\phi}$, and 5 DWT coefficients of $\boldsymbol{\sigma^2}$ need to be incorporated into the wavelet-PAR model and so the final model contains only 13 parameters as opposed to 36 parameters in the original PAR model.
\begin{table}
\tbl{DWT analysis of $\mu$ estimates, $\phi$ estimates and $\sigma^2$ estimates of
$\text{PAR}_{12}(1)$. The statistically significant DWT coefficients ($|Z| > 2.95$) are denoted by *. The coefficients considered in the wavelet-PAR model are indicated
in bold.}
{\begin{tabular}{|c|cccccc|}
\hline
   & \multicolumn{6}{c|}{Parameter}                                                                                                                                                                                                                                                                                              \\ \cline{2-7} 
   & \multicolumn{2}{c|}{$\mu$}                                                                                     & \multicolumn{2}{c|}{$\phi$}                                                                                    & \multicolumn{2}{c|}{$\ sigma^2 $}                                                         \\ \cline{2-7} 
   & \multicolumn{1}{c|}{\begin{tabular}[c]{@{}c@{}}DWT \\ coefficient\end{tabular}} & \multicolumn{1}{c|}{Z-score} & \multicolumn{1}{c|}{\begin{tabular}[c]{@{}c@{}}DWT\\  coefficient\end{tabular}} & \multicolumn{1}{c|}{Z-score} & \multicolumn{1}{c|}{\begin{tabular}[c]{@{}c@{}}DWT\\  coefficient\end{tabular}} & Z-score \\ \hline
0  & \multicolumn{1}{c|}{\textbf{408.88}}                                            & \multicolumn{1}{c|}{-}       & \multicolumn{1}{c|}{\textbf{0.39}}                                              & \multicolumn{1}{c|}{-}       & \multicolumn{1}{c|}{\textbf{2608.33}}                                           & -       \\
1  & \multicolumn{1}{c|}{\textbf{112.69*}}                                           & \multicolumn{1}{c|}{16.91}   & \multicolumn{1}{c|}{-0.19}                                                      & \multicolumn{1}{c|}{-0.95}   & \multicolumn{1}{c|}{\textbf{1393.30*}}                                          & 5.47    \\
2  & \multicolumn{1}{c|}{\textbf{80.35*}}                                            & \multicolumn{1}{c|}{16.96}   & \multicolumn{1}{c|}{-0.02}                                                      & \multicolumn{1}{c|}{-0.12}   & \multicolumn{1}{c|}{\textbf{848.10*}}                                           & 4.60    \\
3  & \multicolumn{1}{c|}{\textbf{-56.61*}}                                           & \multicolumn{1}{c|}{-10.52}  & \multicolumn{1}{c|}{0.07}                                                       & \multicolumn{1}{c|}{0.35}    & \multicolumn{1}{c|}{-603.82}                                                    & -2.93   \\
4  & \multicolumn{1}{c|}{6.75}                                                       & \multicolumn{1}{c|}{1.15}    & \multicolumn{1}{c|}{-0.24}                                                      & \multicolumn{1}{c|}{-1.19}   & \multicolumn{1}{c|}{278.60}                                                     & 1.19    \\
5  & \multicolumn{1}{c|}{\textbf{-69.44*}}                                           & \multicolumn{1}{c|}{-10.87}  & \multicolumn{1}{c|}{-0.37}                                                      & \multicolumn{1}{c|}{-1.87}   & \multicolumn{1}{c|}{-639.87}                                                    & -2.49   \\
6  & \multicolumn{1}{c|}{\textbf{80.40*}}                                            & \multicolumn{1}{c|}{14.36}   & \multicolumn{1}{c|}{-0.46}                                                      & \multicolumn{1}{c|}{-2.32}   & \multicolumn{1}{c|}{\textbf{898.89*}}                                           & 3.99    \\
7  & \multicolumn{1}{c|}{-7.24}                                                      & \multicolumn{1}{c|}{-1.24}   & \multicolumn{1}{c|}{0.05}                                                       & \multicolumn{1}{c|}{0.26}    & \multicolumn{1}{c|}{-469.41}                                                    & -2.01   \\
8  & \multicolumn{1}{c|}{-2.07}                                                      & \multicolumn{1}{c|}{-0.38}   & \multicolumn{1}{c|}{-0.10}                                                      & \multicolumn{1}{c|}{-0.48}   & \multicolumn{1}{c|}{202.58}                                                     & 0.85    \\
9  & \multicolumn{1}{c|}{-0.53}                                                      & \multicolumn{1}{c|}{-0.09}   & \multicolumn{1}{c|}{-0.02}                                                      & \multicolumn{1}{c|}{-0.11}   & \multicolumn{1}{c|}{-69.71}                                                     & -0.29   \\
10 & \multicolumn{1}{c|}{4.87}                                                       & \multicolumn{1}{c|}{0.84}    & \multicolumn{1}{c|}{-0.12}                                                      & \multicolumn{1}{c|}{-0.62}   & \multicolumn{1}{c|}{147.40}                                                     & 0.59    \\
11 & \multicolumn{1}{c|}{\textbf{-75.58*}}                                           & \multicolumn{1}{c|}{-13.48}  & \multicolumn{1}{c|}{0.17}                                                       & \multicolumn{1}{c|}{0.87}    & \multicolumn{1}{c|}{\textbf{-737.15}}                                           & -3.17   \\
12 & \multicolumn{1}{c|}{4.84}                                                       & \multicolumn{1}{c|}{0.93}    & \multicolumn{1}{c|}{-0.16}                                                      & \multicolumn{1}{c|}{-0.79}   & \multicolumn{1}{c|}{143.75}                                                     & 0.64    \\
13 & \multicolumn{1}{c|}{-2.82}                                                      & \multicolumn{1}{c|}{-0.51}   & \multicolumn{1}{c|}{0.21}                                                       & \multicolumn{1}{c|}{1.03}    & \multicolumn{1}{c|}{0.37}                                                       & 0.00    \\
14 & \multicolumn{1}{c|}{1.22}                                                       & \multicolumn{1}{c|}{0.22}    & \multicolumn{1}{c|}{-0.15}                                                      & \multicolumn{1}{c|}{-0.74}   & \multicolumn{1}{c|}{308.65}                                                     & 1.30    \\
15 & \multicolumn{1}{c|}{-10.45}                                                     & \multicolumn{1}{c|}{-1.90}   & \multicolumn{1}{c|}{0.05}                                                       & \multicolumn{1}{c|}{0.25}    & \multicolumn{1}{c|}{-681.49}                                                    & -2.87   \\ \hline
\end{tabular}}
\label{ballypatrick wav resuts}
\end{table}
\begin{table}
\tbl{Box-Pierce test for wavelet-PAR residuals.}
{\begin{tabular}{ccc}
\toprule
Lags    & 20     & 30     \\ \hline
p-value & 0.3159 & 0.4438  \\ \bottomrule
\end{tabular}}
\label{ballypatrick wav box}
\end{table}
\section{Conclusion}
PARMA models are used extensively in various fields like economics, climatology, signal processing and hydrology. However, the presence of a substantial number of parameters in a PARMA model reduces its efficiency. Fourier methods already exist in the literature for reducing the number of parameters and have found some success when the parameters vary slowly. However, transient events are quite common in real-life applications and wavelet techniques stand out as the principal analysis tool in these situations. The efficiency of wavelet methods in reducing the number of parameters of PARMA models has been demonstrated using a simulation study and their applicability in real life has been illustrated using real data. This study opens new avenues for the reduction of parameters in other classes of models containing a large number of parameters.

\section*{Acknowledgement}
Rhea Davis acknowledges the financial support of the University Grants Commission (UGC), India under the Savitribai Jyotirao Phule Fellowship for Single Girl Child (SJSGC) scheme.

\section*{Disclosure statement}
There is no conflict of interest between the authors.

\bibliographystyle{tfs}
\bibliography{main}
\section{Appendices}
In this section, we state the results used to analyse our models, which were proved under the following assumptions:\\
\hfill\\
Let $\{Y_t\}$ be a process defined by (\ref{PARMAeq}). Then,
\begin{enumerate}
\itemB Finite Fourth moment: $E \varepsilon_t^4 = \eta <\infty$.
\itemB The model admits a causal representation
$$
Y_t=\sum_{j=0}^{\infty} \psi_t(j) \varepsilon_{t-j},
$$
where $\psi_t(0)=1$ and $\sum_{j=0}^{\infty}\left|\psi_t(j)\right|<\infty$ for all $t$. Note that $\psi_t(j)=\psi_{t+k \nu}(j)$ for all $j$. Also, $Y_t=\tilde{Y}_t-\mu_t$ and $\varepsilon_t=\sigma_t \delta_t$, where $\{\delta_t\}$ is i.i.d.
\itemB The model satisfies an invertibility condition
$$
\varepsilon_t=\sum_{j=0}^{\infty} \pi_t(j) Y_{t-j},
$$
where $\pi_t(0)=1$ and $\sum_{j=0}^{\infty}\left|\pi_t(j)\right|<\infty$ for all $t$. Again, $\pi_t(j)=\pi_{t+k\nu}(j)$ for all $j$.
\itemB The spectral density matrix $f(\lambda)$ of the equivalent vector ARMA process given by,
\begin{equation}
    U_t = \sum_{j=-\infty}^{\infty}\Psi_jZ_{t-j},
\end{equation}
where $U_t= (Y_{t\nu},\dotsc,Y_{(t+1)\nu-1})^{\prime}$, $Z_t=(\varepsilon_{t\nu},\dotsc,\varepsilon_{(t+1)\nu-1})^{\prime}$ and $\Psi_t$ is the $\nu \times \nu$ matrix with $ij$ entry $\psi_i(t\nu+i-j)$ and for some $0<c \leq C<\infty$ we have,
$$
c z^{\prime} z \leq z^{\prime} f(\lambda) z \leq C z^{\prime} z, \quad-\pi \leq \lambda \leq \pi,
$$
for all $z$ in $R^{\nu}$.
\itemB The number of iterations $n$ of the iterations algorithm satisfies $n \leq N v-1$, and $n^2 / N \rightarrow 0$ as $N \rightarrow \infty$ and $n \rightarrow \infty$.
\itemB The number of iterations $n$ of the iterations algorithm satisfies $n \leq N v-1$, and $n^3 / N \rightarrow 0$ as $N \rightarrow \infty$ and $n \rightarrow \infty$.
\end{enumerate}

\begin{theoremA}[\textbf{Innovation Algorithm for Periodically Stationary Processes,} see \cite{anderson1999innovations}]\label{A.innovationalgstat}
    If $\left\{Y_t\right\}$ has zero mean and $E\left(Y_{\ell} Y_m\right)=\gamma_{\ell}(m-\ell)$, where the ma$\operatorname{trix} \Gamma_{n, i}=\left[\gamma_{i+n-1-\ell}(\ell-m)\right]_{\ell, m=0, \ldots, n-1}, i=0, \ldots, v-1$, is nonsingular for each $n \geqslant 1$, then the one-step predictors $\hat{Y}_{i+n}, n \geqslant 0$, and their mean-square errors $v_{n, i}, n \geqslant 1$, are given by
$$
\hat{Y}_{i+n}= \begin{cases}0 & \text { if } n=0 \\ \sum\limits_{j=1}^n \theta_{n, j}^{(i)}\left(Y_{i+n-j}-\hat{Y}_{i+n-j}\right) & \text { if } n \geqslant 1\end{cases},
$$
and for $k=0,1, \ldots, n-1,$
\begin{center}
    $v_{0, i}=\gamma_i(0),$
\end{center}
\begin{equation}\label{thetainnoeq}
\theta_{n, n-k}^{(i)}=\left(v_{k, i}\right)^{-1}\left[\gamma_{i+k}(n-k)-\sum_{j=0}^{k-1} \theta_{k, k-j}^{(i)} \theta_{n, n-j}^{(i)} v_{j, i}\right],
\end{equation}
\begin{equation}\label{vinnoeq}
v_{n, i}=\gamma_{i+n}(0)-\sum_{j=0}^{n-1}\left(\theta_{n, n-j}^{(i)}\right)^2 v_{j, i}.
\end{equation}
where (\ref{thetainnoeq}) is solved recursively in the order $v_{0, i} ; \theta_{1,1}^{(i)}, v_{1, i} ; \theta_{2,2}^{(i)}, \theta_{2,1}^{(i)}, v_{2, i} ; \theta_{3,3}^{(i)}, \theta_{3,2}^{(i)}, \theta_{3,1}^{(i)}$, $v_{3, i}, \ldots$.\\
The estimates of $\theta_{n,j}^{(i)}$ and $v_{n,i}$, $\hat{\theta}_{n,j}^{(i)}$ and $\hat{v}_{n,i}$ respectively, can be obtained by substituting the estimates of autocovariance function (\ref{gamma.est}). 
\end{theoremA}
\begin{theoremA}[see \cite{anderson1999innovations}]\label{A.psi theorem}
    Let $\tilde{Y}_t$ be a process as defined in (\ref{PARMAeq}). Under the assumptions B1-B5, we have, 
    \begin{equation}
\hat{\theta}_{n,j}^{(\langle i-n \rangle)} \xrightarrow{p} \psi_i(j), \quad \text{as} \, n \rightarrow \infty, \quad \text{for all} \, j,\, i = 0, 1, \dotsc, \nu - 1.
\end{equation}
Here the notation, $\langle b \rangle$ is $b-\nu \lfloor b/{\nu} \rfloor \, \text{for} \, b = 0, 1, \dotsc$ and $\nu+b-\nu\lfloor b/{\nu} + 1 \rfloor$ for $b = -1, -2, \dotsc$, where $\lfloor \cdot \rfloor$ is the greatest integer function.
\end{theoremA}
\begin{theoremA}[see \cite{anderson1999innovations}]\label{A.con.sigma2}
    Let $\tilde{Y}_t$ be a process as defined in (\ref{PARMAeq}). Under the assumptions B1-B5, we have, 
    \begin{equation}
\hat{v}_{n,\langle i-n \rangle} \xrightarrow{p} \sigma_i^2, \quad \text{as} \, n \rightarrow \infty,\, i = 0, 1, \dotsc, \nu - 1.
\end{equation}
Here the notation, $\langle b \rangle$ is $b-\nu \lfloor b/{\nu} \rfloor \, \text{for} \, b = 0, 1, \dotsc$ and $\nu+b-\nu\lfloor b/{\nu} + 1 \rfloor$ for $b = -1, -2, \dotsc$, where $\lfloor \cdot \rfloor$ is the greatest integer function.
\end{theoremA}
\begin{theoremA}[see \cite{tesfaye2011asymptotic}]\label{A.phi.equal.psi.result}
    Let $\{Y_t\}$ be a $\text{PAR}_{\nu}(1)$ as given in (\ref{particularPAR}) having causal representation B2. Then, $\{\phi_0, \phi_1, \dotsc, \phi_{\nu-1}\}^{\prime} = \{\psi_0(1), \psi_1(1), \dotsc, \psi_{\nu-1}\}^{\prime}$. 
\end{theoremA}
\begin{theoremA}[see \cite{anderson2005parameter}, \cite{tesfaye2011asymptotic}]\label{A.asym.dist.phi}
    Let $\{Y_t\}$ be a $\text{PAR}_{\nu}(1)$ as given in (\ref{particularPAR}). Under the assumptions B1-B4 and B6, and using Theorem \ref{A.phi.equal.psi.result}, we have, 
    \begin{equation}\label{asym.distbn.psi}
N^{1 / 2}(\hat{\phi}_{i}-\phi_i) \Rightarrow \mathcal{N}(0, \sigma_{i-1}^{-2}\sigma_i^2), \quad i = 0, 1, \dotsc, \nu - 1.
\end{equation}
\end{theoremA}
\begin{theoremA}[see \cite{anderson2021parsimonious}]\label{A.asymp.distbn.mu}
Let $\tilde{Y}_t$ be a process as defined in (\ref{PARMAeq}) having causal representation B2. Then, 
    \begin{equation}
        N^{1/2}(\boldsymbol{\hat{\mu}}-\boldsymbol{\mu}) \Rightarrow \mathcal{N}(0,\Sigma_{\boldsymbol{\mu}}),
    \end{equation}
   where $\Sigma_{\boldsymbol{\mu}} = \sum_{n = \infty}^{\infty}B_n \Pi^n, B_n = \text{diag} \, (\gamma_0(n),\gamma_1(n), \dotsc, \gamma_{\nu-1}(n)),$ and $\Pi$ is as given in (\ref{Pi}).
\end{theoremA}
\begin{theoremA}[see \cite{anderson2021parsimonious}]\label{A.asymp.distb.var}
    Let $\tilde{Y}_t$ be a process as defined in (\ref{PARMAeq}). Under assumptions B1-B5, and if,
    \begin{equation}
 N^{3/4} \sum_{l>n} |\pi_i(l)| \rightarrow 0 \, \text{as} \, N \rightarrow \infty \, \text{and} \, n \rightarrow \infty,   
\end{equation}we have,
\begin{equation}
    \begin{pmatrix}
\hat{v}_{n,\langle 0-n \rangle}\\ 
\hat{v}_{n,\langle 1-n \rangle}\\ 
\vdots \\ 
\hat{v}_{n,\langle \nu-1-n \rangle}
\end{pmatrix} \, \text{is} \, \text{AN} \left [ \begin{pmatrix}
\sigma_0^2\\ 
\sigma_1^2\\ 
\vdots \\ 
\sigma_{\nu-1}^2
\end{pmatrix} , N^{-1} \begin{pmatrix}
s_{0,0} & \dotsc & s_{0,\nu - 1} \\ 
 \vdots & \cdots  &\vdots \\ 
 s_{\nu - 1,0}&\dotsc  & s_{\nu - 1,\nu - 1}
\end{pmatrix}\right],
\end{equation}
where $s_{i,j}$ is given by,
\begin{equation*}
    s_{i,j} = \sum_{m_1=0}^{\infty} \sum_{m_2=0}^{\infty}\pi_i(m_1)\pi_{j}(m_2)(W_{m_1m_2})_{i-m_1,j-m_2},
\end{equation*}
and $(W_{m_1,m_2})_{i,j}$ is given by,
\begin{align*}
\left(W_{m_1 m_2}\right)_{i,j}= & (\eta-3) \sum_{j_1=-\infty}^{\infty} \psi_i\left(j_1\right) \psi_{i+m_1}\left(j_1+m_1\right) \\
& \cdot \sum_{n=-\infty}^{\infty} \psi_{j}\left(j_1+n \nu+j-i\right) \psi_{j+m_2}\left(j_1+n \nu+j-i+m_2\right) \sigma_{i-j_1}^4 \\
& +\sum_{n=-\infty}^{\infty}\left[\gamma_i(n \nu+j-i) \gamma_{i+m_1}\left(n \nu+j-i-m_1+m_2\right)\right. \\
& \left.+\gamma_i\left(n \nu+j-i+m_2\right) \gamma_{i+m_1}\left(n \nu+j-i-m_1\right)\right], \, \text{for all} 
\end{align*}
with $\eta = E(\delta_t^4)$ and $\{\delta_t\} = \{\sigma_t^{-1}\epsilon_t\}$.
\end{theoremA}
\begin{theoremA}[see \cite{tesfaye2011asymptotic}]\label{A.PARMA.estimates}
    Let $\tilde{Y}_t$ be PARMA(1,1) process as defined in (\ref{particularPARMA}) having causal representation B2. Then, 
    \begin{equation}
    \phi_t = \frac{\psi_t(2)}{\psi_{t-1}(1)}.
\end{equation}
\begin{equation}
    \theta_t = \phi_t - \psi_t(1).
\end{equation}
\end{theoremA}
\begin{theoremA}[see \cite{tesfaye2011asymptotic}]
\label{A.Q.matrix.thm}
    Let $\tilde{Y}_t$ be the PARMA(1,1) process as defined in (\ref{particularPARMA}). Using assumptions B1-B4 and B6, we have, 
        \begin{equation}
        N^{(1/2)}(\hat{\boldsymbol{\phi}} - \boldsymbol{\phi}) \implies \mathcal{N}(0,\mathcal{Q}),
    \end{equation}
where $\boldsymbol{\hat{\phi}} = [\hat{\phi}_0, \dotsc, \hat{\phi}_{\nu-1}]^{\prime}, \boldsymbol{\phi} = [\phi_0, \dotsc, \phi_{\nu-1}]^{\prime}$ and $\nu \times \nu$ matrix $\mathcal{Q}$ is defined by
\begin{equation}
    \mathcal{Q} = \sum_{k,l=1}^{2}\mathcal{H}_{\ell}\mathcal{V}_{\ell k}\mathcal{H}_{k}^{\prime},
\end{equation}
where 
\begin{equation} \label{A.Vlk}
    \mathcal{V}_{\ell k} = \sum_{j=1}^{x} (\{\mathcal{F}_{\ell-j} \Pi^{\ell-j})\mathcal{B}_j(\mathcal{F}_{k-j}\Pi^{-(k-j)})^{\prime}\},\, x = min(\ell, k),
\end{equation}
\begin{equation} 
    \mathcal{F}_j = \text{diag}\{\psi_0(j),\psi_1(j),\dotsc,\psi_{\nu-1}(j)\},
\end{equation}
\begin{equation}
    \mathcal{B}_j = \text{diag}\{\sigma_0^2\sigma_{0-j}^2, \sigma_1^2\sigma_{1-j}^2 \dotsc, \sigma_{\nu-1}^2\sigma_{\nu-1-j}^2\},
\end{equation}
$\mathcal{H}_1 = -\mathcal{F}_2\Pi^{-1}\mathcal{F}_1^{-2}$, $\mathcal{H}_2 = \Pi^{-1}\mathcal{F}_1^{-1}\Pi$ and $\Pi$ is as defined in (\ref{Pi}).
\end{theoremA}
\begin{theoremA}[see \cite{tesfaye2011asymptotic}]\label{A.S.matrix.thm}
    Let $\tilde{Y}_t$ be the PARMA(1,1) process as defined in (\ref{particularPARMA}). Using assumptions B1-B4 and B6, we have,
    \begin{equation}
        N^{(1/2)}(\hat{\boldsymbol{\theta}} - \boldsymbol{\theta}) \implies \mathcal{N}(0,\mathcal{S}),
    \end{equation}
where 
$\boldsymbol{\hat{\theta}} = [\hat{\theta}_0, \dotsc, \hat{\theta}_{\nu-1}]^{\prime}, \boldsymbol{\theta} = [\theta_0, \dotsc, \theta_{\nu-1}]^{\prime}$ and $\nu \times \nu$ matrix $\mathcal{S}$ is defined by
\begin{equation}
    \mathcal{S} = \sum_{k,l=1}^{2}\mathcal{M}_{\ell}\mathcal{V}_{\ell k}\mathcal{M}_{k}^{\prime},
\end{equation}
where 
$\mathcal{M}_1 = \mathcal{I} - \mathcal{F}_2\Pi^{-1}\mathcal{F}_1^{-2}$, $\mathcal{M}_2 = \Pi^{-1}\mathcal{F}_1^{-1}\Pi,$ $\mathcal{V}_{\ell k}$ is given in (\ref{A.Vlk}) and $\Pi$ is as given in (\ref{Pi}). Here, $\mathcal{I}$ is the $\nu \times \nu$ identity matrix.
\end{theoremA}
\begin{theoremA}[see \cite{anderson2021parsimonious}]\label{A.LPU.rep}
    For a given vector $\mathbf{X}$, the periodic function $\mathbf{X} = [X_t: 0 \leq t \leq \nu - 1]$ has the Fourier representation
\begin{equation} 
    X_t = c_{0}+\sum_{r=1}^{\ell}\left\{c_{r}cos\left(\frac{2\pi rt}{\nu}\right)+s_{r}sin\left(\frac{2\pi r t}{\nu}\right)\right\},
\end{equation}
where $c_{r}$ and $s_{r}$ are Fourier coefficients and 
\begin{equation*}
    \ell = \begin{cases}
        \nu /2 \quad \text{if} \, \nu \, even \\
        (\nu -1)/2 \quad \text{if}\,  \nu \, odd
    \end{cases}.
\end{equation*}
Let $f = \begin{cases}
    [c_{0},c_{1},s_{1},\dotsc,c_{(\nu-1)/2},s_{(\nu-1)/2}]^{\prime} \quad (\nu \, \text{odd}) \\
    [c_{0},c_{1},s_{1},\dotsc,s_{(\nu /2 -1)},c_{(\nu /2)}]^{\prime} \quad (\nu \, \text{even})
\end{cases}$. 
\begin{equation}
\text{Then,} \quad f = \mathcal{L}\mathcal{P}\mathcal{U}\mathbf{X},
\end{equation}
where $\mathcal{L}$ and $\mathcal{U}$ are respectively given by, 
\begin{equation}\label{A.L.exp}
    \mathcal{L} = \begin{cases}
        \text{diag} (\nu^{-1/2}, \sqrt{2/ \nu}, \dotsc, \sqrt{2/ \nu} ) \quad \quad \quad \quad (\nu \, \text{odd}) \\
        \text{diag} (\nu^{-1/2}, \sqrt{2/ \nu}, \dotsc, \sqrt{2/ \nu}, \nu^{-1/2} ) \quad (\nu \, \text{even})
    \end{cases}, 
\end{equation}
and 
\begin{equation}\label{A.U.exp}
    \mathcal{U} = \nu^{-1/2}\left(e^{\frac{-i2\pi rt}{\nu}}\right).
\end{equation}
The $(\ell,j)$th element of the $\mathcal{P}$ is given by,
\begin{equation}\label{A.P.exp}
    [\mathcal{P}]_{\ell j}= \begin{cases}1 & \text { if } \ell=j=0 ; \\ 2^{-1 / 2} & \text { if } \ell=2 r-1 \text { and } j=r \text { for some } 1 \leq r \leq[(v-1) / 2] \\ 2^{-1 / 2} & \text { if } \ell=2 r-1 \text { and } j=v-r \text { for some } 1 \leq k \leq\lfloor(v-1) / 2\rfloor \\ i 2^{-1 / 2} & \text { if } \ell=2 r \text { and } j=r \text { for some } 1 \leq r \leq[(v-1) / 2] ; \\ -i 2^{-1 / 2} & \text { if } \ell=2 r \text { and } j=v-r \text { for some } 1 \leq k \leq\lfloor(v-1) / 2\rfloor ; \\ 1 & \text { if } \ell=v-1 \text { and } j=v / 2 \text { and } v \text { is even; and } \\ 0 & \text { otherwise. }\end{cases}.
\end{equation}
\end{theoremA}
\begin{theoremA}[see \cite{brockwell1991time}]
    \label{A.broc.asymnormaltheorem}
If $\mathbf{X}_n$ is a $k \times 1$ vector satisfying,
\begin{equation}
    \mathbf{X}_n \Rightarrow \mathcal{N}(\boldsymbol{\mu_n},\Sigma_n),
\end{equation}
and $\mathcal{B}$ is any non-zero $m \times k$ matrix such that the matrices $\mathcal{B}\Sigma_n \mathcal{B}^{\prime}$ have no zero diagonal elements then, 
\begin{equation}
    \mathcal{B}\mathbf{X}_n \Rightarrow \mathcal{N}(\mathcal{B}\boldsymbol{\mu}_n,\mathcal{B}\Sigma_n \mathcal{B}^{\prime}).
\end{equation}
\end{theoremA}
\end{document}